\documentclass[showpacs,preprintnumbers,amsmath,amssymb]{revtex4}

\usepackage{graphicx}
\usepackage{bm}

\def\sq{{I\hskip-2pt I}}
\begin{document}
\title{How to produce quantum entanglement for ascertaining
incompatible properties in double-slit experiments}
\author{G. Nistic\`o, A. Sestito}
\affiliation{Istituto Nazionale Fisica Nucleare, Italy and\\
Dipartimento di Matematica, Universit\`a della Calabria, via P.
Bucci 30b, 87036, Rende (Italy)\\ email: gnistico@unical.it,
sestito@mat.unical.it}

\begin{abstract} Double-slit experiment very well lends itself in
describing the problem of measuring simultaneously incompatible
properties. In such a context, we theoretically design an ideal
experiment for spin-7/2 particles, able to produce the entanglement
which makes possible the detection.
\end{abstract}
\pacs{03.65.Ca, 03.65.Db, 03.65.Ta}

 \maketitle

\section{Introduction}
In Quantum Mechanics the algebraic structure of the set of
observables is mirrored in the algebraic structure of the
self-adjoint operators, which is not commutative; then, whereas in
Classical Mechanics all the observables can be measured together on
a physical system, this is not the case in Quantum Mechanics
\cite{1}. Here such a possibility is restricted to those observables
corresponding to commutative self-adjoint operators.\par The
double-slit experiment is a very effective example in describing
such a phenomenon: the incompatibility between the position
observables at different times makes impossible to measure
simultaneously which slit each particle passes through (WS property)
and the localization on the final screen ($F(\Delta)$). This
notwithstanding, over the years a lot of devices have been conceived
in order to reach ``indirect" knowledge about WS property by
measuring a different property, $T$, correlated with it \cite{2,3}.
More recently, experiments have been proposed  which can give
indirect knowledge about either WS property or an incompatible one,
according to their set-up \cite{9, 10}. The theoretical
investigation performed in \cite{16} in the context of double-slit
experiment, establishes that, under suitable  hypothesis and
exploiting entanglement, the question whether experimental
situations may be conceived, which make possible to obtain
``indirect" knowledge about two incompatible properties (WS and an
incompatible one, G), is positively  answered, in spite of the
predicted impossibility of measuring them simultaneously. Following
such a result, in \cite{17} an ideal double-slit experiment for
atoms is designed such that the values of two non-commuting quantum
properties are inferred simultaneously by revealing photons emitted
by single atoms within micro-maser cavities.\par
 Other approaches in literature face the problem of ascertaining simultaneously  incompatible
 properties. The claimed impossibility \cite{11} of producing inferences
 for more than three observables urges to consider the situation
 of detecting two incompatible properties, besides WS, together with
 the measurement of the final impact point. The theoretical
 investigation in \cite{18} shows that also this question has
 affirmative answer within this approach; an ideal experiment
 realizing such a detection is presented in \cite{17}. \par
 In the present work, after introducing the problem the simultaneous detection
 of three incompatible properties, WS, G and L,
 together with the final impact point  (section 2), an ideal double slit experiment is designed
 which realizes such a detection (section 3). We notice that
 this ideal experiment correspond to a particular solution of the
  theoretical investigation in \cite{18} and with the assumptions
  therein, properties $L$ and $G$ turn out always correlated.
  Moreover, it must be
  stressed that such a detection requires an
  entanglement between the particle and the detector.  Section 3 is
  devoted to show how such an entanglement can be ideally realized.
\section{Detecting incompatible properties simultaneously}
 We consider a system (e.g. the nucleus of an atom) which can travel towards the
 two slits, but not elsewhere, whose position is represented, in
 Heisenberg picture, by an operator in the Hilbert space ${\cal
 H}_I$. It possesses further degrees of freedom (e.g. spin) described
 in the Hilbert-space ${\cal H}_\sq$; hence,
${\mathcal H}={\mathcal H}_I\otimes{\mathcal H}_\sq$ is the Hilbert
space describing the entire system. In general, we denote by $A_I$
($A_\sq$) an operator of ${\cal H}_I$ (${\cal H}_\sq$). The
projection operator representing WS property ``the particle passes
through slit 1 (2)" has the form $E=E_I\otimes{\bf 1}_\sq $
($E'=({\bf 1}_{I}-E_I)\otimes{\bf 1}_\sq $). Given any interval
$\Delta$ on the final screen, we denote the projection operator
representing the property ``the particle hits the final screen in a
point within $\Delta$" by $F(\Delta)$. Though $E$ and $F(\Delta)$
are both localization projections, they refers to different times --
the time $t_1$ of the slits' crossing for $E$ and the time $t_2>t_1$
of the final impact for $F(\Delta)$. We suppose that the Hamiltonian
operator in independent of the degrees of freedom described by
${\mathcal H}_\sq$, so that it has the form  $H=H_I\otimes {\bf
1}_\sq $; hence,  it can be shown that $F(\Delta)=F_I(\Delta)\otimes
{\bf 1}_\sq$ follows, but $[E,F(\Delta)]\neq{\bf 0}$ must hold too
\cite{18}. Thus, it is not generally possible to ascertain WS
property and the final impact point, by  direct localization
measurements. Rather than measuring $E$, a property
 $T={\bf 1}_{I}\otimes T_\sq$ acting on
${\cal H}_\sq$ can be measured together with $F(\Delta)$, whose
outcomes are correlated with the outcome of $E$,  as expressed by
the following general definition of \emph{detector}
\cite{14}:\\
{\bf Definition 1. }\emph{A projection operator $S$ of ${\cal H}$ is
called a detector of a property $R$ 
with respect to a state $\psi$ if\\
\begin{tabular}{l}
(i) $[S,F(\Delta)]={\bf 0}$,\hspace{30pt} (ii) $[S,R]={\bf 0}$ and
$S\psi=R\psi$
\end{tabular}}.\\[10pt]
Condition (i) ensures that $S$ can be measured together with
$F(\Delta)$; condition (ii) allows us to infer the outcome of $R$
(albeit not measured) from the outcome of $S$ \cite{4}. According to
Def. 1, if for a given state $\Psi$, a projection operator $T={\bf
1}_{I}\otimes T_\sq$ exists such that $T\Psi=E\Psi$, then it is
possible to \emph{detect} which slit each particle passes through by
means of a measurement of $T$; indeed, since  $E=E_I\otimes {\bf
1}_\sq$ and $F(\Delta)=F_I\otimes {\bf 1}_\sq$, conditions
$[T,F(\Delta)]={\bf 0}$ and $[T,E]={\bf 0}$ are automatically
satisfied. Hence, in other words, outcome $1$ ($0$) for $T$ reveals
the passage of the particle through slit $1$ ($2$).
\par We seek for the possibility of detecting two further properties,
$G$ and $L$: let $G=G_I\otimes {\bf 1}_\sq$ and $L=L_I\otimes {\bf
1}_\sq$ be other properties, incompatible with each other and  with
WS property $E$, i.e $[L,G]\neq 0$, $[L,E]\neq 0$ and $[E,G]\neq 0$;
if for a given state $\Psi$, two commuting projection operators
$Y={\bf 1}_{I}\otimes Y_\sq$ and $W={\bf 1}_{I}\otimes W_\sq$ exist
such that $Y$ is a detector for $G$ and $W$ is a detector for $L$,
then the outcomes of $Y$ and $W$ reveal the occurrence of the
properties $G$ and $L$ respectively; therefore, the problem is:

\par\vspace{10pt} \noindent {\sl Problem}
(${\cal P'}$). \emph{Given WS property $E=E_{I}\otimes {\bf 1}_\sq $
we have to find
\begin{itemize}
\item[-]  two projection operators $G_{I}$ and $L_{I}$ of ${\cal H}_I$,
\item[-]  three projection operators $T_\sq $, $Y_\sq $ and $W_\sq $ of ${\cal H}_\sq $,
\item[-] a state vector $\Psi\in {\cal H}_I\otimes {\cal H}_\sq $,
\end{itemize}
such that the following conditions are satisfied:\\
\begin{tabular}{lll}
      (C.1) $[E,G]\neq 0$ i.e $[E_{I},G_{I}]\neq 0\qquad$ &
            (C.2) $[E,L]\neq 0$ i.e $[E_{I},L_{I}]\neq 0,$\cr
      (C.3) $[L,G]\neq 0$ i.e $[L_{I},G_{I}]\neq 0,\qquad$&
            (C.4) $[T,E]= 0$ and $T\Psi= E\Psi$\cr 
      (C.5) $[Y,G]= 0$ and $Y\Psi= G\Psi \qquad$& 
            (C.6) $[W,L]= 0$ and $W\Psi= L\Psi$\cr 
      (C.7) $[T,Y]= 0$& 
            (C.8) $[T,W]= 0$\cr 
      (C.9) $[Y,W]=0$& 
            (C.10) $\Psi\neq E\Psi\neq 0$, $\Psi\neq G\Psi\neq 0$ and $\Psi\neq L\Psi\neq 0$. 
\end{tabular}}\par\vspace{10pt}
The  ideal experiment  described in the next section represents a
solution of problem (${\cal P'}$). \vskip2cm

\section{An Ideal Experiment for simultaneous detection on incompatible properties}
The system of our ideal experiment is a spin-7/2 particle whose
position observable is described in a Hilbert space $\mathcal{H}_I$,
while the spin observables are described in
$\mathcal{H}_\sq\equiv{\bf C}^8$. Let $\psi_i$, (resp.,
$\psi_{i+5}$), $i=1,...,5$ be 5 mutually orthonormal vectors of
$\mathcal{H}_I$ localized in slit 1 (resp., slit 2) when the
particle crosses the slits' support, i.e. such that
$E_I\psi_i=\psi_i$ (resp., $E_I\psi_{i+5}=0$). No further condition
is required to these vectors. These ten vectors form an orthonormal
set. Then we take the Hilbert space $\mathcal{H}_I$ as the space
generated by them. This implies that
\begin{equation}\protect\label{eq:E_I}
E_I\varphi=\langle\psi_1\mid\varphi\rangle_I\psi_1+
\langle\psi_2\mid\varphi\rangle_I\psi_2+
\langle\psi_3\mid\varphi\rangle_I\psi_3+
\langle\psi_4\mid\varphi\rangle_I\psi_4+
\langle\psi_5\mid\varphi\rangle_I\psi_5\quad\hbox{for every }\;
\varphi\in\mathcal H_I\,.
\end{equation}
There are 8 eigenvectors $\alpha_1=\vert 7/2\rangle$,
$\alpha_2=\vert 5/2\rangle$, $\alpha_3=\vert 3/2\rangle$,
$\alpha_4=\vert 1/2\rangle$, $\alpha_5=\vert -1/2\rangle$,
$\alpha_6=\vert -3/2\rangle$, $\alpha_7=\vert -5/2\rangle$,
$\alpha_8=\vert -7/2\rangle \in\mathcal{H}_\sq$ corresponding to the
8 possible values (in $\hbar$ units) of the spin along direction
$z$, represented by the hermitian operator $S_z$ of ${\bf C}^8$.
\par
Let the particle be prepared in the entangled state represented by
\begin{eqnarray}\protect\label{eq:Psi}
\Psi&=&
\frac{1}{32}\left(-\psi_1-2\psi_2+\psi_3+\psi_4+\psi_5\right)\vert
7/2\rangle+\nonumber\\
&&+\frac{1}{8}\sqrt\frac{7}{10}\left(-\psi_1+\psi_2-2\psi_3+3\psi_5\right)
\vert 3/2\rangle +\frac{\sqrt{
35}}{32}\left(-\psi_6-2\psi_7+\psi_8+\psi_9+\psi_{10}\right)\vert
1/2\rangle+\nonumber\\
&&+\frac{1}{16}\left[\sqrt\frac{35}{11}\left(\psi_1+\psi_2+3\psi_4\right)+
\left(4\psi_1+\psi_2+3\psi_3+3\psi_5\right)\right]\vert-1/2\rangle+\nonumber\\
&&+\frac{1}{8}\sqrt\frac{7}{30}\left(-\psi_6+\psi_7-2\psi_8+3\psi_9\right)\vert-5/2\rangle+\nonumber\\
&&+\frac{1}{16}\left[\frac{1}{\sqrt{11}}\left(\psi_6+\psi_7+3\psi_9\right)+
\frac{1}{\sqrt{35}}\left(4\psi_6+\psi_7+3\psi_8+3\psi_{10}\right)\right]\vert-7/2\rangle.
\end{eqnarray}
The 8 projection operators $A^i_\sq= \vert j_i\rangle\langle
j_i\vert$, $i=1,...,8$  represent spin observables pertaining
$\mathcal{H}_\sq$ (if $A={\bf 1}_I\otimes A^i_\sq$ has outcome 1,
then the particle has spin $j_i=\frac{7}{2} - (i - 1)$" along $z$
and so on). They trivially commute with both $F(\Delta)$ and $E$.
Then also the projection operator $T=A^1+A^2+A^3+A^5={\bf
1}_I\otimes(
\vert7/2\rangle\langle7/2\vert+\vert5/2\rangle\langle5/2\vert+\vert3/2\rangle\langle/2\vert+\vert-1/2\rangle\langle-1/2\vert)$
commute with $F(\Delta)$ and $E$. Now, straightforward calculations
show that $E\Psi=T\Psi$. 
Therefore, $T$ turns out to be a WS detector.
\par
Now we introduce a property $G=G_I\otimes {\bf 1}_\sq $ incompatible
with $E$, which can be detected by means of a suitable detector $Y$
without renouncing to the WS knowledge provided by $T$. Given any
$\varphi\in\mathcal H_I$, we define
\begin{equation}\protect\label{eq:G_I}G_I\varphi=\langle\psi^{(1)}\mid\varphi\rangle_I\psi^{(1)}+
\langle\psi^{(2)}\mid\varphi\rangle_I\psi^{(2)}+
\langle\psi^{(3)}\mid\varphi\rangle_I\psi^{(3)}
\end{equation} where
\begin{eqnarray*}&\psi^{(1)}&=\frac{1}{6}\psi_1-\frac{1}{6}\psi_2-\frac{1}{6}\psi_1-\frac{\sqrt
3}{2}\psi_7+\frac{1}{2\sqrt 3}\psi_9+\frac{1}{2\sqrt 3}\psi_{10}\\
&\psi^{(2)}&=-\frac{\sqrt 3}{2}\psi_1+\frac{1}{2\sqrt
3}\psi_2+\frac{1}{2\sqrt 3}\psi_{3}-\frac{\sqrt
6}{4}\psi_6+\frac{\sqrt 6}{4}\psi_8+\frac{1}{2\sqrt
6}\psi_9+\frac{1}{2\sqrt 6}\psi_{10}\\
&\psi^{(3)}&=-\frac{\sqrt 2}{4}\psi_1-\frac{\sqrt
2}{2}\psi_2+\frac{\sqrt 2}{4}\psi_3+\frac{\sqrt
2}{4}\psi_4+\frac{\sqrt 2}{4}\psi_5.
\end{eqnarray*}
\par\noindent A straightforward calculation based on
(\protect\ref{eq:E_I}) and (\protect\ref{eq:G_I}) shows that
$[G_I,E_I]\varphi\neq 0$ so that $G$ and $E$ are incompatible with
each other. However, the projection operator $Y=A^1+A^2+A^4+A^6={\bf
1}_I\otimes(
\vert7/2\rangle\langle7/2\vert+\vert5/2\rangle\langle5/2\vert+\vert1/2\rangle\langle1/2\vert+\vert-3/2\rangle\langle-3/2\vert)$
satisfies condition $Y\Psi=G\Psi$ and it trivially commutes with
$F(\Delta)=F_I(\Delta)\otimes{\bf 1}_\sq$; therefore $Y$ is a
detector of $G$.\par Now we introduce a further property
$L=L_I\otimes {\bf 1}_\sq $ incompatible with $E$ and $G$, which can
be detected by means of a suitable detector $W$ without renouncing
to the WS knowledge provided by $T$ and to the knowledge of $G$
provided by $L$. Given any $\varphi\in\mathcal H_I$, we define
\begin{equation}\protect\label{eq:L_I}L_I\varphi=\langle\psi^{[1]}\mid\varphi\rangle_I\psi^{[1]}+
\langle\psi^{[2]}\mid\varphi\rangle_I\psi^{[2]}+
\langle\psi^{[3]}\mid\varphi\rangle_I\psi^{[3]}+
\langle\psi^{[4]}\mid\varphi\rangle_I\psi^{[4]}+\langle\psi^{[5]}\mid\varphi\rangle_I\psi^{[5]}
\end{equation} where
\begin{eqnarray*}&\psi^{[1]}&=\frac{1}{5}\sqrt{\frac{11}{15}}\left(-2\psi_1+2\psi_2+2\psi_3-
                            \frac{9}{\sqrt{11}}\psi_6+\frac{9}{\sqrt{11}}\psi_7+\frac{9}{\sqrt{11}}\psi_{10}
                              \right)\\
                 &\psi^{[2]}&=\frac{1}{10}\sqrt{\frac{11}{65}}\left(3\psi_1-3\psi_2-3\psi_3
                              -\frac{24}{\sqrt{11}}\psi_6-\frac{51}{\sqrt{11}}\psi_7
                              +\frac{25}{\sqrt {11}}\psi_9+\frac{49}{\sqrt {11}}\psi_{10}\right)\\
                 &\psi^{[3]}&=\frac{1}{5}\sqrt{\frac{33}{26}}\left(-\psi_1+\psi_2+\psi_3-
                              \frac{17}{6\sqrt{11}}\psi_6-\frac{14}{3\sqrt{11}}\psi_7 +
                              \frac{65}{6\sqrt{11}}\psi_8+\frac{5}{2\sqrt{11}}\psi_9-
                              \frac{11}{2\sqrt{11}}\psi_{10}\right)\\
                 &\psi^{[4]}&=\frac{1}{\sqrt{15}}\left(-\psi_1+\psi_2-2\psi_3+3\psi_5\right)\\
                 &\psi^{[5]}&=\frac{1}{2\sqrt 2}\left(-\psi_1-2\psi_2+\psi_3+\psi_4+\psi_5\right).\\
\end{eqnarray*}
\par\noindent A straightforward calculation
shows that $[G_I,L_I]\varphi\neq 0$ so that $G$ and $L$ are
incompatible with each other; furthermore, $[E_I,L_I]\varphi\neq 0$
so that $E$ and $L$ are incompatible with each other, too. However,
the projection operator $W=A^1+A^3+A^4+A^7={\bf 1}_I\otimes(
\vert7/2\rangle\langle7/2\vert+\vert3/2\rangle\langle3/2\vert+\vert1/2\rangle\langle1/2\vert+\vert-5/2\rangle\langle-5/2\vert)$
satisfies condition $W\Psi=L\Psi$ and it trivially commutes with
$F(\Delta)=F_I(\Delta)\otimes{\bf 1}_\sq$; therefore $W$ is a
detector of $L$.\par Nevertheless, we have that $W$, $Y$ and $T$
pairwise commute with each other. Then all, $W$, $Y$ and $T$ can be
simultaneously measured together with the position of the final
impact; in other words,  properties $L$, $G$ and $E$, incompatible
with each other, can be detected together on each particle localized
on the final screen. In the present work, we are not concerned with
the question of the physical meaning of $G$ and $L$, which evidently
depends upon  the choice of  vectors $\psi_i$, $\psi_{i+5}$,
$i=1,...,5$. Thus, we have a solution of problem ($\mathcal P'$).

\section{Experimental issues for entanglement} In the perspective of a realization
of the  detection of $E$, $G$ and $L$, a crucial experimental task
is to create the entanglement, encoded in the state vector $\Psi$ in
(\protect\ref{eq:Psi}), between the particle and the detectors,
before the time $t_1$ when the particle reaches the screen
supporting the slits. This can be (ideally) realized in three steps.
In the first step only particles with the $x$ component of the spin
equal to 7/2 are selected, for instance by means of a suitable
Stern-Gerlach apparatus. Hence, in Schroedinger picture, at this
stage - time $t_0<t_1$ - the state vector is of the kind $\psi\vert
s\rangle$, with $\psi\in{\mathcal H}_I$, $\Vert\psi\Vert=1$, and
$S_x\vert s\rangle=7/2\vert s\rangle$, so that we can take $\vert
s\rangle=\frac{1}{8\sqrt2} \{\vert 7/2\rangle+\sqrt 7\vert
5/2\rangle+\sqrt{21}\vert3/2\rangle+\sqrt{35}\vert
1/2\rangle+\sqrt{35}\vert
-1/2\rangle+\sqrt{21}\vert-3/2\rangle+\sqrt 7\vert -5/2\rangle+\vert
-7/2\rangle\}$.
\par In the second step, during their flight between times $t_0$ and
$t_1$, the particles undergo the action of another Stern-Gerlach
magnet, able to spatially separate the particles with respect to
$S_z$: the particles with $S_z=7/2$, $5/2$, $3/2$ or $-1/2$ (resp.
$-7/2$, $-5/2$, $-3/2$ or $1/2$) are forced to travel towards slit 1
(resp. 2), but through two alternative spatial channels according to
the value of $S_z$, so that the dynamical evolution between times
$t_0$ and  a time $t_{1/2}<t_1$ is represented by a unitary operator
$U$ such that
\begin{eqnarray*}
&U(\psi\vert 7/2\rangle)=\psi_1^{[\frac{7}{2}]}\vert
7/2\rangle,\qquad & U(\psi\vert
-7/2\rangle)=\psi_2^{[-\frac{7}{2}]}\vert -7/2\rangle,\\
&U(\psi\vert 5/2\rangle)=\psi_1^{[\frac{5}{2}]}\vert
5/2\rangle,\qquad &U(\psi\vert
-5/2\rangle)=\psi_2^{[-\frac{5}{2}]}\vert -5/2\rangle,\\
&U(\psi\vert 3/2\rangle)=\psi_1^{[\frac{3}{2}]}\vert
3/2\rangle,\qquad & U(\psi\vert
-3/2\rangle)=\psi_2^{[-\frac{3}{2}]}\vert -3/2\rangle,\\
&U(\psi\vert -1/2\rangle)=\psi_1^{[-\frac{1}{2}]}\vert
-1/2\rangle,\qquad & U(\psi\vert
1/2\rangle)=\psi_2^{[\frac{1}{2}]}\vert 1/2\rangle.
\end{eqnarray*}
where $\psi_k^{[J]}$ are vectors of ${\mathcal H}_I$ representing
the alternative spatial channels taken by the particles to reach
slit $k$; hence $\langle \psi_{k_1}^{[J_1]
}\mid\psi_{k_2}^{[J_2]}\rangle=\delta_{k_1,k_2}\cdot\delta_{J_1,J_2}$
and $E_I\psi_1^J=\psi_1^{[J]}$, $E_I\psi_2^{[J]}=0$. Then the
outcoming state must be $\widehat{\Psi}=U(\psi\vert
s\rangle)=\sum_{j_i,k}^{} \psi_k^{[j_i]}\vert j_i\rangle$.\par
Before reaching the slits, between the times $t_{1/2}$, $t_1$, the
particles undergo the action of a filter blocking the beams of
particles corresponding to $S_z=5/2$ and  $-3/2$: the state
$\widehat{\Psi}$ is of the kind $\Psi_0+\Psi_1$, with
$\Psi_0=\psi_1^{[\frac{5}{2}]}\vert
5/2\rangle+\psi_2^{[-\frac{3}{2}]}\vert -3/2\rangle$ and
$\Psi_1=\sum_{j_i\neq 5/2, -3/2,k}^{} \psi_k^{[j_i]}\vert
j_i\rangle$; the effect of the filter is represented by the
projection operator $P=\vert \Psi_1\rangle\langle \Psi_1\vert$ so
that the final state vector outcoming the whole preparing procedure
is
\begin{equation*}\Psi=P\widehat{\Psi}=\frac{1}{8\sqrt2}
\{\psi_1^{[\frac{7}{2}]}\vert
7/2\rangle+\sqrt{21}\psi_1^{[\frac{3}{2}]}\vert3/2\rangle+\sqrt{35}\psi_1^{[\frac{1}{2}]}\vert
1/2\rangle+\sqrt{35}\psi_1^{[-\frac{1}{2}]}\vert -1/2\rangle+\sqrt
7\psi_1^{[-\frac{5}{2}]}\vert
-5/2\rangle+\psi_1^{[-\frac{7}{2}]}\vert
-7/2\rangle\}.\end{equation*}\par
Now, if dim$(E_I{\mathcal H}_I)$,
dim$(({\bf 1}_I-E_I){\mathcal H}_I)\geq 5$, then five mutually
orthonormal vectors $\psi^i\in E_I{\mathcal H}_I$,
$\psi^{i+5}\in({\bf 1}_I- E_I){\mathcal H}_I$, with $i=1,\ldots,5$,
exist such that
\begin{eqnarray*}
&\psi_1^{[\frac{7}{2}]}&=\frac{1}{2\sqrt
2}\left(-\psi_1-2\psi_2+\psi_3+\psi_4+\psi_5\right),\\
&\psi_1^{[\frac{3}{2}]}&=\frac{1}{\sqrt
{15}}\left(-\psi_1+\psi_2-2\psi_3+3\psi_5\right),\\
&\psi_2^{[\frac{1}{2}]}&=\frac{1}{2\sqrt
2}\left(-\psi_6-2\psi_7+\psi_8+\psi_9+\psi_{10}\right),\\
\end{eqnarray*}
\begin{eqnarray*}
&\psi_1^{[-\frac{1}{2}]}&=\frac{1}{\sqrt
{22}}\left(\psi_1+\psi_2+3\psi_4\right)+\frac{1}{\sqrt
{70}}\left(4\psi_1+\psi_2+3\psi_3+3\psi_5\right),\\
&\psi_2^{[-\frac{5}{2}]}&=\frac{1}{\sqrt
{15}}\left(-\psi_6+\psi_7-2\psi_8+3\psi_9\right),\\
&\psi_2^{[-\frac{7}{2}]}&=\frac{1}{\sqrt
{22}}\left(\psi_6+\psi_7+3\psi_9\right)+\frac{1}{\sqrt
{70}}\left(4\psi_6+\psi_7+3\psi_8+3\psi_{10}\right),
\end{eqnarray*}
then the state $\Psi$ outcoming from this dynamical preparing
process must be
\begin{eqnarray*}
\Psi&=&P(U(\psi \vert s\rangle=
\frac{1}{32}\left(-\psi_1-2\psi_2+\psi_3+\psi_4+\psi_5\right)\vert
7/2\rangle+\\
&&+\frac{1}{8}\sqrt\frac{7}{10}\left(-\psi_1+\psi_2-2\psi_3+3\psi_5\right)
\vert 3/2\rangle
+\frac{\sqrt{35}}{32}\left(-\psi_6-2\psi_7+\psi_8+\psi_9+\psi_{10}\right)\vert
1/2\rangle+\\
&&+\frac{1}{16}\left[\sqrt\frac{35}{11}\left(\psi_1+\psi_2+3\psi_4\right)+
\left(4\psi_1+\psi_2+3\psi_3+3\psi_5\right)\right]\vert-1/2\rangle+\\
&&+\frac{1}{8}\sqrt\frac{7}{30}\left(-\psi_6+\psi_7-2\psi_8+3\psi_9\right)\vert-5/2\rangle+\\
&&+\frac{1}{16}\left[\frac{1}{\sqrt{11}}\left(\psi_6+\psi_7+3\psi_9\right)+
\frac{1}{\sqrt{35}}\left(4\psi_6+\psi_7+3\psi_8+3\psi_{10}\right)\right]\vert-7/2\rangle,
\end{eqnarray*}
which is just the state vector (\protect\ref{eq:Psi}) which carries
the right entanglement allowing for a simultaneous detection  of the
mutually incompatible properties $L$, $G$, $E$, together with the
measurement of the final impact point.

\end{document}